\documentclass[acmsmall]{acmart}
\AtBeginDocument{%
  \providecommand\BibTeX{{%
    \normalfont B\kern-0.5em{\scshape i\kern-0.25em b}\kern-0.8em\TeX}}}

\setcopyright{acmcopyright}
\copyrightyear{2021}
\acmYear{2021}
\acmDOI{10.1145/1122445.1122456}

\acmBooktitle{ToCHI}
\acmPrice{15.00}
\acmISBN{978-1-4503-XXXX-X/18/06}

\usepackage{xurl}
\usepackage{xspace} 

\usepackage{tabularx}
\usepackage{booktabs}
\newlength{\toprulewidth}
\usepackage{multirow}
\usepackage{subcaption}

\newcommand{\numberofparticipants}{227\xspace}

\begin{document}

\title{Remote VR Studies}
\subtitle{A Framework for Running Virtual Reality Studies Remotely Via Participant-Owned HMDs}

\author{Rivu Radiah}
\email{sheikh.rivu@unibw.de}
\affiliation{%
  \institution{Bundeswehr University Munich}
  \country{Germany}
}

\author{Ville Mäkelä}
\email{ville.maekelae@ifi.lmu.de}
\orcid{0000-0001-6095-2570}
\affiliation{%
  \institution{LMU Munich}
  \country{Germany}
}

\author{Sarah Prange}
\email{sarah.prange@unibw.de}
\orcid{0000-0001-8303-1600}
\affiliation{%
  \institution{Bundeswehr University Munich}
  \country{Germany}
}

\author{Sarah Delgado Rodriguez}
\email{sarah.delgado@unibw.de}
\affiliation{%
  \institution{Bundeswehr University Munich}
  \city{Munich}
  \country{Germany}
}

\author{Robin Piening}
\email{robin.piening@unibw.de}
\affiliation{%
  \institution{LMU Munich}
  \country{Germany}
}

\author{Yumeng Zhou}
\affiliation{%
  \institution{LMU Munich}
  \country{Germany}
}

\author{Kay Köhle}
\affiliation{%
  \institution{LMU Munich}
  \country{Germany}
}

\author{Ken Pfeuffer}
\email{ken.pfeuffer@unibw.de}
\affiliation{%
  \institution{Bundeswehr University Munich}
  \city{Munich}
  \country{Germany}
}

\author{Yomna Abdelrahman}
\email{yomna.abdelrahman@unibw.de}
\affiliation{%
  \institution{Bundeswehr University Munich}
  \country{Germany}
}

\author{Matthias Hoppe}
\email{matthias.hoppe@ifi.lmu.de}
\affiliation{%
  \institution{LMU Munich}
  \country{Germany}
}

\author{Albrecht Schmidt}
\email{albrecht.schmidt@ifi.lmu.de}
\orcid{0000-0003-3890-1990}
\affiliation{
\institution{LMU Munich}
  \country{Germany}
}

\author{Florian Alt}
\email{florian.alt@unibw.de}
\orcid{0000-0001-8354-2195}
\affiliation{%
  \institution{Bundeswehr University Munich}
  \country{Germany}
}

\renewcommand{\shortauthors}{Rivu et al.}

\begin{abstract}
We investigate opportunities and challenges of running virtual reality (VR) studies remotely. Today, many consumers own head-mounted displays (HMDs), allowing them to participate in scientific studies from their homes using their own equipment. Researchers can benefit from this approach by being able to reach a more diverse study population and to conduct research at times when it is difficult to get people into the lab (cf. the COVID pandemic). We first conducted an online survey (N=\numberofparticipants), assessing HMD owners' demographics, their VR setups and their attitudes towards remote participation. We then identified different approaches to running remote studies and conducted two case studies for an in-depth understanding. We synthesize our findings into a framework for remote VR studies, discuss the strengths and weaknesses of the different approaches, and derive best practices. Our work is valuable for HCI researchers conducting VR studies outside labs.
\end{abstract}



\maketitle

\section{Introduction}

Virtual Reality (VR) has become a widely adopted technology in the Human-Computer Interaction (HCI) research community and beyond  \cite{zhang2016survey}. Many subtopics emerged around this technology, including but not limited to,  novel interaction techniques, presence and immersion, avatar modeling, navigation, and locomotion. More recently, researchers started looking into how VR can substitute or complement other research approaches \cite{Maekelae:2020:VirtualFieldStudies,Evaluating}, for example, in-the-wild studies that require a lot of effort to maintain the technology used, evaluations in potentially dangerous environments (e.g., automotive or pedestrian user interfaces), or the investigation of situations that occur rarely. 

In the past decades, VR headsets were not widely available and their setup required a substantial degree of technical knowledge. Hence, the vast majority of research was conducted in lab settings, where researchers took care of the setup and guided participants through (controlled) studies. As a result, participants of such studies were often part of a rather homogeneous university population -- a well-known challenge in HCI research \cite{10.1145/1240624.2180963}.

The past years witnessed a proliferation of VR head-mounted displays (HMDs) in the consumer market\footnote{Consumer VR market size: \url{https://www.statista.com/statistics/528779/virtual-reality-market-size-worldwide/}} and while the number of worldwide shipped VR units with 6 million units in 2019\footnote{Shipment of VR devices by vendor: \url{https://www.statista.com/statistics/671403/global-virtual-reality-device-shipments-by-vendor/}} is still marginal from a commercial perspective, a considerable number of end users now own a VR headset. This creates an unprecedented chance for researchers to move VR research out of the lab and reach a more diverse community -- similar to what happened with other technologies that reached the consumer market, such as smartphones  \cite{Henze10,henze2011,Henze11niels} or public displays \cite{davies2014pervasive}.

In this work, we contribute an in-depth investigation of an emerging research paradigm, that is conducting out-of-the-lab research with owners of VR headsets. In particular, we explore the challenges and pitfalls researchers are facing as they are shifting or complementing their research using this new paradigm and synthesize them into a framework for use by other researchers. To this end, our work is guided by the following research questions:

\begin{itemize}
    \item \textbf{RQ1:} How suitable are HMD owners and their VR setups as subjects for remote studies, and what limitations do home setups have?
    \begin{itemize}
        \item \textbf{1.1:} What are the demographics of HMD owners?
        \item \textbf{1.2:} Which VR equipment do they own?
        \item \textbf{1.3:} What are their home VR setups like?
        \item \textbf{1.4:} How do they feel about participating in remote studies from home?
    \end{itemize}
    \item \textbf{RQ2:} What suitable approaches exist for conducting remote VR studies, and what are their advantages and disadvantages?
    \begin{itemize}
        \item \textbf{2.1:} What options exist for developing and setting up VR research applications?
        \item \textbf{2.2:} How can VR research prototypes be distributed to study participants?
        \item \textbf{2.3:} Through which channels can owners of VR headsets be reached and recruited?
        \item \textbf{2.4:} What special considerations are there when designing and running a remote study?
    \end{itemize}
\end{itemize}




Our research approach is as follows: we set out with conducting an online survey (N=\numberofparticipants) among owners of VR headsets to understand their demographics, the equipment they own, the setting in which they use it, and whether they would be willing to participate in remote VR studies.  Subsequently, based on a review of existing VR distribution channels, we describe different ways in which VR applications for remote studies can be implemented and distributed as well as how data collection can be realized. We then report on our experiences from two remote case studies, where we experimented with different approaches. We then synthesize all our findings into a framework that guides researchers through different ways of designing and conducting remote VR studies.

\noindent Our framework consists of four primary approaches to remote studies:
\begin{enumerate}
    \item  Researchers build a standalone VR application that is distributed directly to participants.
    \item Researchers build a VR application that is distributed through existing vendor platforms ("app stores") such as Steam and the Play Store.
    \item  Researchers build a VR application using an API of an existing social VR platform (e.g., Rec Room, VRChat) and upload it to the corresponding platform.
    \item Researchers set up their study environment directly within an existing social VR platform using the tools provided by these platforms.
\end{enumerate}

All of these approaches come with unique advantages, drawbacks, and considerations. Some of the clearest differences are in the level of autonomy and the amount of required effort. Using existing vendor platforms and social VR platforms requires that researchers conform to the regulations and limitations of these platforms. This often means that options for data collection are limited, and the platforms may also be more limited in their functionalities. However, using such platforms can often make setting up the study far easier and faster, and existing platforms ensure an existing user base, and often alleviate compatibility issues. Building custom software and distributing it independently may get around many of the limitations with existing platforms, but the workload is often higher and may result in other challenges (e.g., ensuring the safety and privacy of participants).

Another dimension is whether or not the VR study is conducted with a remote study experimenter, or whether participants run the study independently, i.e., without remote guidance. We discuss how well the four main approaches lend themselves to guided studies and independent studies.



Our research is valuable for researchers considering to conduct virtual reality studies remotely. Beyond an opportunity to reach a larger and more diverse audience, our research is also of value for other reasons. Most prominently, the recent COVID-19 pandemic affected the HCI community around the globe, making it difficult for many researchers to recruit participants for their work in the lab. This unprecedented event raised the question: "How can VR researchers continue to conduct studies?" Here, we believe that our work can provide useful guidance. Our findings are not only valuable in a pandemic but also opens up opportunities beyond these times. 

\vspace{1mm}\noindent\textbf{Contribution Statement.} We contribute a holistic view on remote VR studies. Specifically, (1) we present the results of an online survey assessing users' view towards such studies and their VR settings; (2) we provide a review of research approaches with a focus on how they influence the development / study setup, software distribution, and recruitment challenges; (3) we report on two studies and derive lessons learned; and (4) we synthesize our findings into a framework. 


\section{Background} 



Several strands of prior work are relevant to our research. After briefly introducing the term \emph{remote VR studies}, we review (a) work involving ubiquitous technologies for conducting out-of-the-lab research as well as (b) prior work on conducting VR studies outside the lab.

\subsection{Terminology}

Over many decades, the HCI research community adapted or came up with new approaches to research, which can be broadly classified into research conducted in controlled settings (i.e. in the lab) vs. in less controlled settings (i.e. in the field). As Virtual Reality is being appropriated as a tool for  conducting research, another dimension -- in addition to where the study physically takes place -- emerges, that is the virtual setting. Here, researchers are provided rich opportunities to situate their research in almost any setting they like -- be it the reconstruction of a lab, a public space, a replica of a person's home, or an entire world (cf. Second Life\footnote{Second Life: \url{https://secondlife.com}, last accessed January 21, 2021}). 

We focus on studies taking place physically in users' homes but not being restricted with regard to the virtual setting. In the following, we refer to these type of studies as \emph{remote VR studies}.


\subsection{Leveraging Consumer Devices for Out-of-the-Lab Research}

The idea of leveraging consumers' devices for 'bringing the lab to participants' is already established practice in several areas of human-computer interaction research. Among the most prominent examples is work by Henze et al. who used smart phones as a tool to conduct out-of-the-lab studies, initially on touch targeting and typing behavior \cite{Henze10,henze2011,Henze11niels}. Over the following years, the community saw more studies focusing on various aspects, including but not limited to an exploration of user interaction with notifications \cite{Sahami2014CHI}, authentication behavior \cite{alt2015mobilehci,buschek2016snapapp,schneegass2014ubicomp}, and keystroke dynamics  \cite{buschek2018researchime}.


Prior work has identified both strength and challenges of such types of studies. Of particular interest is the work of Gustarini et al. \cite{Gustarini13} who categorize challenges into the design, development, execution, and data analysis phase. While their work is specific to studies focusing on smart phones, several of the challenges apply to other technologies as well and, hence, informed our work. These challenges include the question of which data can be collected, where data can be stored, how researchers can deal with device heterogeneity, recruiting bias, remunerating participants, participant cheating, handling participants' questions, considering participants' motivation and privacy needs, conducting interviews and how to synchronize data.

Studies have also been conducted to understand the extent of ecological validity in remote experiments. For example, Andreasen et al. \cite{Andreasen07} conducted an empirical comparison of remote usability testing and conventional lab testing, and Germine et al. \cite{germine2012web} investigated data quality across lab-based and web-based experiments. Research shows that it is possible to achieve ecologically valid results using out-of-lab studies \cite{Andreasen07,germine2012web}.

\subsection{Virtual Reality Out-of-the-Lab Studies}

In the following section  we summarize key insights obtained from prior research. We take a chronological view on how out-of-the-lab studies have progressed over the years.

In 2012, Hodgson el at. presented ways for `portable' VR studies, using systems that can be easily carried and used in-the-wild \cite{hodgson12}. The authors introduced a self-contained backpack, capable of running VR simulations and demonstrated how such portable systems can be used by researchers.

Four years later, Steed et al. \cite{Steed16} experimented on presence and embodiment in VR through remote studies by recruiting Samsung Gear VR and Google Cardboard owners. They argue that for remote VR studies it is generally easier to recruit more participants than for typical lab-based studies. However, their work suggests that the diversity of participants was still defined by the nature of the study. The authors also highlighted that remote studies might require more effort than lab studies in terms of, for example, preparation and app development.

In 2017, Mottelson et al. \cite{Mottelson:2017:VirtualRealityStudies} conducted experiments both in and out of the lab to understand the advantages and pitfalls between the two approaches. The experiments generally yielded comparable results as both in-lab and remote studies were able to collect reliable data and similar output. At the same time, out-of-the-lab experiments were characterized by a greater heterogeneity among participants such as sample population. They also outline several limitations of remote studies including demographics validity, ethical concerns, and the lack of experiment control.

Ma et al. \cite{Xiao18} conducted three behavioral experiments in VR using crowd sourcing in 2018. Their evaluation suggests the feasibility of conducting web-based VR experiments albeit several challenges, including few participants with access to VR-capable devices. The authors also highlight many advantages, including obtaining a more diverse sample compared to lab VR studies.  In a study about social interactions in weekly VR get-togethers, Moustafa and Steed \cite{Moustafa18} in the same year also highlighted the potential of reaching broad and diverse audiences through remote studies. 

Recently, Saffo et al. investigated the feasibility of running VR studies via a popular social platform, VRChat \cite{Saffo:2020:CrowdsourcingVR}. They identified the ability to enable researchers to create their own content, the large user base, and the ease of recruiting participants from within VRChat as advantages of the platform as it is easier to recruit participants within the platform. At the same time, Saffo et al. reported on data collection to be limited when using online platforms as each online platform reviews the content before researchers are able to upload content. 

Huber and Gajos \cite{Huber:2020:ConductingOnlineVirtualExperiments} worked on conducting online VR studies with uncompensated and unsupervised participants. While they identified advantages, such as recruiting participants from all over the world with various backgrounds, there are challenges: for example, in this study the sample count was much smaller compared to other online studies. 



\subsection{Summary}

Our review shows that while researchers have recognized and acknowledged the potential of remote VR studies, for many years they have struggled with challenges, such as the available technology, the number of people who had access to the technology, means to reach out to them, and ways of delivering software to the participants. 
Not only with the current pandemic, we see that interest in this type of studies is increasing at a rapidly accelerating pace. We believe to be currently at a turning point where with advances in technology and due to the increasing proliferation of the technology for the first time it is finally possible for VR researchers to strongly benefit from this approach. Yet there is still a lack of a comprehensive understanding of how such studies can be implemented. This is at the focus of our work. 



    





While much can be learned from other sub fields of HCI where technology that became widely available to end users was leveraged for conducting remote studies, there are many aspects unique to virtual reality and head-mounted displays that remain under-explored, including but not limited to the available technology, the environment, data collection methods, and distribution channels.

Some of the aforementioned aspects have been touched upon in prior work. Yet, our analysis shows that pre-requisites have changed over the years and new opportunities and challenges have emerged. Our work provides an understanding of the current state-of-the-art in running remote VR studies as well as a comprehensive assessment of this study paradigm that we synthesize into a framework. The framework explains different approaches to remote VR studies and also summarizes best practices and lessons learned.




\section{Understanding VR Users through an Online Survey}

We conducted an online survey among VR users with the goal of better understanding (a) their demographics, (b) how they can be reached and recruited, (c) what their VR setups look like, (d) which technology they own, and (e) whether they are willing to participate in remote VR studies.

\subsection{Survey Content and Questions}

In particular, our survey focuses on the following aspects:

\begin{description}
    \item[Demographics] We assessed the participants' age, gender, background, and their reasons for owning a VR device.
    \item[VR Equipment \& Platforms] We asked about the devices they own and the platforms they are using. 
    \item[VR Setting and Use] We asked them to describe their VR setup, the space (with an option to upload a picture of the space), how often they use it, and how likely they are to be interrupted while being in VR.
    \item[Willingness to participate]  We asked the participants how willing they are to participate in remote VR studies, both in general as well as in situations such as the COVID-19 lockdown, and what kind of payment they might expect.
\end{description}

%


\subsection{Recruiting}

We distributed the survey via different channels, in particular such that would also be used by researchers to reach out to potential participants of a VR study. To this end, we included research-oriented crowd sourcing platforms, VR online forums, social media and university mailing lists.  In the following we report some lessons learned. 

\subsubsection{Research-oriented Crowd Sourcing Platforms}

We recruited through two platforms: Prolific\footnote{Prolific: \url{https://www.prolific.co/}, last accessed January 21, 2021} and XRDRN\footnote{XRDRN: \url{https://www.xrdrn.org/}, last accessed January 21, 2021}. Prolific is a platform primarily meant for recruiting study participants, for example, for surveys. It allows participants to be pre-selected by demographics (for example, owners of VR devices). XRDRN is a dedicated platform to connect researchers and participants with a specific focus on mixed reality. The platform was built specifically in response to the COVID-19 pandemic. Response rates from Prolific were considerably higher and faster. This was likely because (a) Prolific integrates means for payment directly with the platform whereas for XRDRN this would be handled out of the platform between researchers and participants, and (b) XRDRN is still in the early stages of building a community. Another aspect is that since all participants on Prolific are paid, many of them can be expected to be `professional' study takers. 

\subsubsection{Forums \& Social Media}
We reached out to VR users through Facebook, Reddit and platform-specific forums (e.g., Steam, Rec Room). Within these platforms, we identified suitable subgroups (e.g., VR-related subreddits in Reddit).

Within platforms, subgroups could be easily identified. However, it took some time to ensure that we complied with the rules of each subgroup and that posting an invitation to an online survey was appropriate. Where groups did not specifically allow or disallow for these kinds of posts, we reached out to moderators to seek permission. While for some groups we obtained permission in this way, many moderators were unsure and suggested we rather post in other channels, for example, such that were concerned with the organization of VR events or for discussing miscellaneous topics. We noticed that such groups often had fewer active members, which limited the visibility of the survey.

Some subgroups employed specific posting procedures. Some required specific tags or features to be assigned to messages (for example, on Reddit or Rec Room) -- yet the pre-defined tags were often a poor match for our recruitment messages. Many forums also make it mandatory for users to achieve a certain number of points (points are acquired based on several metrics such as number of comments and posts on other discussion threads) before they are eligible to create their own threads, while others required a user account to be at least two months old before being eligible to post. When posting links to different subforums simultaneously, there is a chance that auto-moderators flag the posts as spam. 
Established VR products (such as Unity and Rec Room) run forums and groups on different social platforms (such as Reddit, Facebook, Twitter, Discord). Those could be used in addition to more general groups on VR.


\subsubsection{University Mailing Lists}

This method helped to reach out to a large number of recipients. In particular, we used mailing lists allowing a general university population to be reached. This population included many non-technical subjects. Our experience shows that only few of them were eligible for the survey (as they were required to own an HMD to participate). Mailing lists targeting students from computer science subjects might reach more participants.

\subsection{Results} 

We received $\numberofparticipants$ complete surveys (out of 276). Respondents were recruited through Prolific (97), Reddit (95), mailing lists (14), Facebook (12), Discord (4), VR platforms (3) and other (2). 


\subsubsection{Demographics}

A total of $\numberofparticipants$ participants responded to our online survey, $52$ female (cf. Table~\ref{tab:online_survey_demographics} for an overview) 
Participants were between $18$ and $56$ years of age ($Mean=29.72, SD=8.8$). They mainly lived in the US ($N=66$), the UK ($N=54$), and Germany ($N=24$).

Most of the respondents identified as \textit{consumers} of VR products ($N=214$, 94\%), and some were VR developers ($N=21$, 9\%), VR content creators ($N=20$, 9\%), and VR researchers ($N=16$, 7\%). They mainly used their VR setup for \textit{gaming} ($N=206$, 91\%), but also for other purposes like socialising, development and research, and during work.

We furthermore asked the participants if they had discovered new uses for their VR setup during the COVID-19 pandemic. More than half ($N=130$, 57\%) reported that they had discovered new VR games to play, and some had discovered new VR platforms ($N=35$, 15\%). Some users had started using their setups for entirely new purposes, like meeting with friends and relatives ($N=24$, 11\%), watching movies socially ($N=23$, 10\%), and collaborating with colleagues ($N=12$, 5\%).

\begin{table}[!t]
    \caption{Online survey: participants' demographics including their relationship to VR, which VR platforms they use (left), their purposes of using VR, and possible new uses discovered during the pandemic (multiple choice questions each) (right).}
    \label{tab:online_survey_demographics}
\parbox {.5\linewidth}{ 
    \centering
    \scriptsize
        \begin{tabularx}{0.8\linewidth}{c|Xr}
    \toprule
    \multirow{4}{*}{\rotatebox[origin=c]{90}{\textit{Gender}}}
    & Male & 170\\
    & Female & 52\\
    & Prefer not to say & 1\\
    & Other & 4\\
    \midrule
     \multirow{4}{*}{\rotatebox[origin=c]{90}{\textit{Roles}}}
    & I am a consumer of VR products and applications. & 214\\
    & I am a developer of VR products and applications & 21\\
    & I create content for VR. & 20\\
    & I conduct research in the area of VR. & 16\\
    & Other & 3\\
    \midrule
     \multirow{12}{*}{\rotatebox[origin=c]{90}{\textit{Used VR Platforms}}}
    & VRChat & 91\\
    & None & 77\\
    & Rec Room & 57\\
    & AltspaceVR & 26\\
    & Mozilla Hubs & 23\\
    & Other & 12\\
    & Bigscreen VR & 8\\
    & Hologate VR & 7\\
    & Cluster & 5\\
    & Playstation VR & 5\\
    & The Wild & 4\\
    & A Township Tale & 2\\
    \bottomrule 
    \end{tabularx} 
    } 
    \hfill
\parbox[t][][t]{.49\linewidth}{ 
\centering
    \scriptsize
        \begin{tabularx}{0.8\linewidth}{c|Xr}
    \toprule
    \multirow{6}{*}{\rotatebox[origin=c]{90}{\textit{Reasons for Using VR}}}
    & Gaming & 206\\
    & Social platforms & 63\\
    & Socializing with friends and family & 43\\
    & Development or Research & 33\\
    & Work (e.g meetings, presentation) & 17\\
    & Other & 14\\
    & Watch Videos (e.g. Netflix or films) & 10\\
    \\
    \midrule
    \multirow{6}{*}{\rotatebox[origin=c]{90}{\textit{Uses Discovered during Pandemic}}}
    & I  discovered new VR games. & 130\\
    & I  discovered new virtual social platforms. & 35\\
    & I  started VR meetings with friends and relatives. & 24\\
    & I  started  watching VR movies together with friends. & 23\\
    & My usage did not change. & 18\\
    & I  started VR collaboration with colleagues. & 12\\
    & Other & 11\\
    \\
    \bottomrule
    \end{tabularx}

} 
\end{table}

\subsubsection{Understanding Users' VR Settings} 

We first asked participants about the \emph{devices they owned}. Most participants used Oculus devices ($N=79$), followed by Valve Index ($N=59$), PlayStation VR ($N=41$), and HTC Vive ($N=23$). 

\begin{figure}
\begin{subfigure}{0.45\textwidth}
  \includegraphics[width=\textwidth]{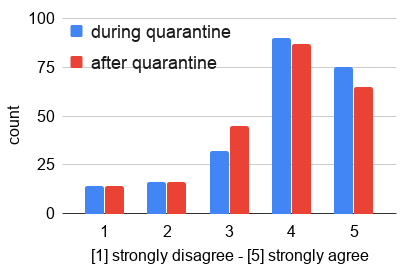}
  \caption{Willingness to participate in studies}
  \label{participation}
\end{subfigure}
\begin{subfigure}{0.45\textwidth}
  \includegraphics[width=\textwidth]{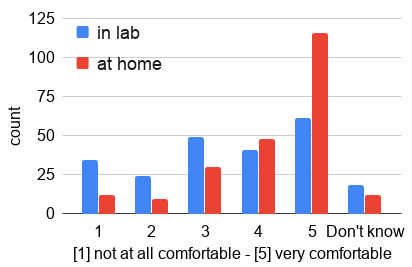}
  \caption{Comfort in participating in studies}
  \label{confidence}
\end{subfigure}
\begin{subfigure}{0.45\textwidth}
  \includegraphics[width=\textwidth]{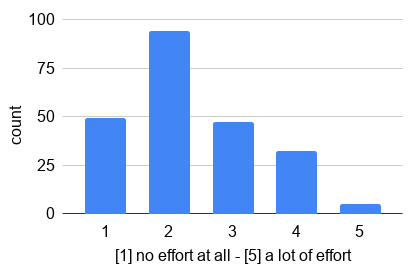}
  \caption{Required effort for VR setup }
  \label{setupeffort}
  \vspace{-2mm}
\end{subfigure}
\caption{Results on how willing participants are to participate in remote studies, how comfortable they feel about it, and how much effort it is for them to use their VR setups.}
\end{figure}

Most participants were the sole users of their VR setup ($N=153$, 67\%). Their setups were placed in rooms rarely used by others ($Median=2$, 1=not at all, 5=almost all the time). Consequently, most participants could use their VR setups without any notable constraints ($N=198$), as opposed to, for example, not being able to access the VR setup due to someone else using the room. Most participants also evaluated that the chances of being interrupted during VR use were low ($Median=2$, 1=very unlikely, 5=very likely).
Furthermore, participants reported that getting their VR setup ready for use (Figure~\ref{setupeffort}) did not require much effort ($Median=2$, 1=no effort at all, 5=a lot of effort).

\begin{figure}[t!]
\begin{subfigure}{0.48\textwidth}
\includegraphics[width=1.0\textwidth]{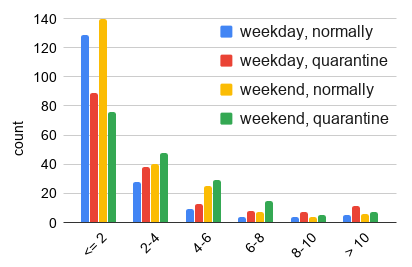}
\caption{Total time (hours) spent in VR on weekdays and weekends. Most users spend at most two hours with the VR setup on weekdays and on weekends under normal circumstances. During the pandemic, participants reported somewhat higher usage.}\label{fig:usage_times}
\end{subfigure}
\hfill
\begin{subfigure}{0.48\textwidth}
\includegraphics[width=1.0\textwidth]{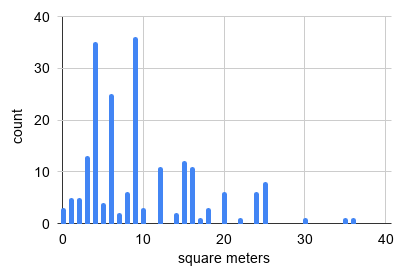}
\caption{Space available for users' VR setups, showing most users have a space of up to 25 square meters. 26 participants stated a larger than 40 square meters space (excluded in this plot).}\label{fig:space_size}
\end{subfigure}
\caption{Results on how much time participants spend in VR and how large their VR setups are.}
\end{figure}

Regarding usage times  (Figure~\ref{fig:usage_times}), most participants reported to normally use their setups for up to two hours on weekdays (in total during an average week)  ($N=129$) as well as up to two hours on weekends ($N=140$) (in total during an average weekend). During the pandemic, participants reported spending overall more time with their VR setup.

%

We asked participants to additionally give an estimate of their VR setup's \emph{size}, providing the width and length in meters (Figure~\ref{fig:space_size}).
65 participants (29\%) reported having up to 5\,$m^2$, 71 (31\%) from 5 to 10\,$m^2$, 26 (11\%) from 10 to 15\,$m^2$, 21 (9\%) from 15 to 20\,$m^2$, and 15 (7\%) from 20 to 25\,$m^2$.

\subsubsection{Participation in VR Studies}

We inquired about participants' \emph{willingness to participate in VR studies} (`I am willing to participate in VR studies.'; 5-Point Likert scale; 1=strongly disagree, 5=strongly agree). To better understand whether this was a result of the \emph{current pandemic situation}, we asked the question for both during and after the lockdown (Figure \ref{participation}). 152 participants (67\%) agreed or strongly agreed to be willing to participate in VR studies after the pandemic ($Median=4$), while 165 participants (73\%) agreed or strongly agreed during the pandemic ($Median=4$).


We also looked into how \emph{comfortable} participants felt about participating in VR studies in different study locations (Figure \ref{confidence}). Participants felt more comfortable about the idea of participating from home than attending a lab study (home: $Median=5$, lab: $Median=3$).

Regarding \emph{compensation} (multiple choices), participants were mostly willing to accept cash ($N=189$, 83\%), but also game vouchers ($N=103$, 45\%). A total of 77 participants (34\%) stated that they would do so voluntarily without any compensation.











\subsection{Summary and Discussion}

Our survey provided insights on the current mainstream users of VR. First, the survey confirms the `cliche' of VR users being primarily gamers (91\%). Furthermore, users were predominantly male (77\%) and mostly consumers (94\%). Besides gaming, a rather prominent use case for VR was socializing, with 28\% of participants stating they use social platforms and 19\% that they use VR to meet with friends and family. We also found that participants used many different VR devices and platforms. The most popular \emph{VR platforms} were VRChat (40\%) and Rec Room (25\%) followed by AltspaceVR (11\%) and Mozilla Hubs (10\%), indicating that there is a wide range of platforms with potential for researchers to conduct studies. Many participants used more than one platform.

Most HMD owners used their VR setups exclusively without notable constraints. Participants reported that the effort required to get their setup ready for use was low, suggesting that VR setups are typically permanent setups with a dedicated space. Respondents also estimated that the likelihood of being interrupted during VR use was low, again suggesting that the setups are mainly used in personal spaces (e.g., bedrooms) rather than in shared spaces (e.g., living rooms). These findings indicate that home VR setups are generally well suited for remote studies.

Furthermore, we assessed the size of the users' VR setups. The majority of participants (60\%) seemed to operate withing a space of up to 10\,$m^2$. Around half of this group (29\% of all respondents) reported having a small space of up to 5\,$m^2$, likely indicating a seated setup. These findings suggest that remote studies requiring active movement in a larger space (e.g., more than 15\,$m^2$) might be challenging as many HMD owners do not have the necessary space, potentially risking skewed research results as well as participants getting injured. To work around the problem of small spaces, researchers might consider various locomotion techniques, such as teleportation techniques \cite{Bozgeyikli:2016:PointAndTeleport,Funk:2019:PointAndTeleport}, to reduce the need for physical movement. However, such techniques might not always be desirable, particularly if "real" physical movement is critical to the phenomena being studied.


The clear majority of participants expressed willingness to participate in remote VR studies (67\% under normal conditions and 73\% during the pandemic). Participants also felt very comfortable about the idea of participating in studies from their own homes, noticeably more so than participating in lab studies. The survey attempts to assess the willingness of a potential participant to participate. In the end, participation is an amalgam of many factors and is dependent on each specific study, such as the participants actually signing up for the study and participating on it, the setup requirements and data collection methods based on the specific research question. Thus, our obtained results indicate an initial interest to participate from home. Our survey also indicates that participants are comfortable about remote participation and they have VR setups which are likely to be permanent, private and uninterrupted during participation. We believe these initial findings are beneficial when researchers are planning remote studies.

Our results also suggest that the COVID-19 pandemic affected people in that they reported more VR usage, and some reported having started to use their VR setups in new ways, like meetings with friends and family. This suggests that users are open to new uses of VR, further strengthening our belief that HMD owners are well willing to participate in remote studies.

We distributed the survey over numerous channels to limit a potential recruiting bias. Nevertheless, we acknowledge that it is hard to fully avoid a recruitment bias. This may have impacted the survey results depending on the popularity of each channel chosen, as some VR platforms may have been more/less popular.

Our results also indicate that HMD owners are highly suitable to be recruited for remote studies. Most respondents were willing to participate in remote studies and felt comfortable about it. Their VR setups and use also support this, as most setups seem to be permanent and used exclusively, with a low chance of interruptions. The most notable challenge identified through our results is that most participants have relatively small setups, which might be challenging for some studies.

\section{Potential Approaches to Remote Studies} 

In this section, we discuss potential ways to conduct remote VR studies and hypothesize about their strengths and weaknesses. We also provide a brief discussion on app stores and social VR platforms that could be utilized for remote VR studies.

\subsection{Development of VR Applications}

\subsubsection{Standalone Applications}

For most VR research projects, researchers develop a custom application that runs independently. This often involves designing and modeling a digital environment, programming the desired functionality, and designing and programming the methods for data collection \cite{Maekelae:2020:VirtualFieldStudies}. The advantage of custom-built applications it that they provide maximum flexibility and freedom when it comes to designing the visuals, means for interaction, data collection, and data storage, among other aspects.

Many modern tools are of great help with building VR applications. Particularly popular in this regard is the Unity engine~\footnote{https://unity.com/, last accessed January 21, 2021}, complemented with the SteamVR plugin~\footnote{https://assetstore.unity.com/packages/tools/integration/steamvr-plugin-32647, last accessed January 21, 2021}. The Unity engine is a massively popular tool for creating games, but it is also widely used for industrial purposes, films, and architecture, among others. As a result of its popularity, the internet is full of tutorials and tips for Unity developers, and communities for supporting developers are plentiful and active. The SteamVR plugin takes care of the basic communication with the head-mounted display and the VR application -- many HMDs, like the HTC Vive, support SteamVR. This greatly alleviates the hurdles related to developing for HMDs. For the most part, developing VR applications is not any different from general 3D development.

Despite these great tools and related resources at our disposal, it is clear that building VR applications takes time and requires considerable software engineering expertise, and in many cases, also expertise in other areas like visual design and 3D modeling. We also anticipate further challenges with remote studies, as researchers have less control over the study and are more in the dark with how participants behave. Researchers may therefore need to develop additional tools to monitor participants or consider other ways to ensure smooth procedures.

\subsubsection{Social VR Platforms}

Another, emerging possibility for remote studies is to use social VR platforms. Many of these platforms, like Rec Room and VRChat, offer diverse customization options that might be enough for many studies. There are considerable differences between such VR platforms in terms of their user base and what features they provide. Based on our analysis of these platforms, we see two distinct approaches to implementing virtual environments for user studies.

First, some VR platforms allow \emph{custom applications to be uploaded}. For example, VR applications built with the Unity engine can be uploaded to VRChat, as long as the applications use the VRChat Software Development Kit (SDK). Publishing applications in VR platforms requires that the application fulfills requirements and follows rules put forth by the platform. This may result in more restricted possibilities as opposed to fully independent VR applications. At the same time, these platforms are likely to bring various advantages. For example, installing and running the application is likely less error-prone and more novice-friendly, and compatibility of different hardware is handled by the platform rather than the application. Furthermore, it might be possible to attract study participants from the existing user base of the platform.

Second, many VR platforms allow users to \emph{build their own environments inside the platform} using simple built-in tools, and offer customization options. This includes choosing between various indoor and outdoor settings and adding 3D objects like furniture, interactive objects like drawing boards, and even questionnaires.
Although it is clear that these tools offer less flexibility and freedom for researchers than fully custom-built applications, these tools make building VR environments considerably faster and easier. Another significant advantage is that these tools are more accessible to researchers from other fields who might have less technical expertise.

According to our survey results, the most popular VR platforms among HMD owners were VRChat, Rec Room, and AltspaceVR. Furthermore, based on our analysis of the features of various VR platforms (Table \ref{tab:vr_platforms_comparison}), these same platforms stand out from the others as they offer the highest flexibility for researchers. Hence, we provide a brief overview of these three platforms. Other VR platforms that are suitable for research purposes may emerge in the future.

\begin{table*}[!t]
  \centering
 \scriptsize
    \begin{tabularx}{\linewidth}{XXXXXX}
    \toprule
    \textbf{Platform} & \textbf{Interactions} & \textbf{Data Collection}  & \textbf{Customizable Avatars} &\textbf{External Sources} & \textbf{Viability}\\
    \midrule
    {Altspace} 
    &Programmable
    &Full (web-hosted)
    &Yes (in-game)
    &Yes (browser)
    &High
    \\
    \midrule
    {VRChat}
    &Programmable
    &Limited
    &Yes (model upload)
    &Yes (Youtube player \& browser)
    &High
    \\
    \midrule
    {Rec Room}
    &In-game rules
    &Limited
    &Yes (model upload)
    &No
    &High
    \\
    \midrule
    {Mozilla Hubs}
    &No Influence
    &No
    &Yes (in-game customisation or model upload)
    &Yes (share media from PC or web)
    &Medium
    \\
    \midrule
    {The Wild}
    &No influence
    &\textit{Insufficient data}
    &No
    &Revit metadata
    &Low
    \\
    \midrule
    {EonReality}
    &Lesson rules
    &\textit{Insufficient data}
    &No avatar
    &Models/Media
    &Low
    \\
    \midrule
    {Minecraft}
    &In-game scripts (Command Blocks)
    &No
    &Yes (reskin)
    &No
    &Low
    \\
    \bottomrule
    \end{tabularx}
    \caption{Comparison of social VR platforms.}
    \vspace{-4mm}
    \label{tab:vr_platforms_comparison}
\end{table*}

\begin{description}

\item[Rec Room] is a popular platform with more than 1 million VR users \cite{Recroomusers}. It allows users to play and create VR games and it is compatible with several operating systems and devices. Users can program simple logic in their custom rooms \cite{Circuits}, making Rec Room a powerful platform for studies that only need simple interactive features.


\item[VRChat] is a highly customizable platform. Custom worlds can be uploaded into the platform (e.g., worlds built using Unity and the VRChatSDK), offering many features (e.g., spawn points, interactive objects). 
Prior research used VRChat for studies \cite{Saffo:2020:CrowdsourcingVR}, finding that this was generally a useful approach that, however, came with several limitations. For example, developers with new accounts cannot upload content without earning a certain 'trust' level. Furthermore, there are compatibility issues with some devices and custom content cannot communicate with outside services.

\item[AltspaceVR] is a social VR platform supporting several devices (e.g, HTC Vive, Oculus). 
It allows new world templates and world objects to be created via Unity Uploader, which can be uploaded and shared on the official hub. Due to security reasons, no scripts are allowed\footnote{\url{https://help.altvr.com/hc/en-us/articles/360015560614-Unity-Uploader-FAQ acc:15.09.2020}}. Developers can, however, place so-called "extensions" into their world through external web pages that get translated into 3D content.

\end{description}

\subsection{Distribution of VR Applications}

In addition to developing and setting up their VR applications, researchers need to think about how the applications will be distributed to participants as well as how potential participants could be reached.

\subsubsection{Direct Download}

Perhaps the most straightforward option is to distribute the application directly to participants, for example, by sending them a download link. Participants then install the application on their own.

It is likely that this approach is relatively effortless and offers the most flexibility, as the application does not need to conform to the rules and regulations of any external services. However, some challenges are likely present. For example, direct installation requires users to install software from untrusted sources (which might limit their willingness to participate) and it might generally be more error-prone than other ways of distribution (e.g., due to lack of support during installation).

\subsubsection{Social VR Platforms}

As already discussed, social VR platforms might be utilized for remote studies, as study environments can be set up through them. Since the VR application resides within an existing platform, there is no need to explicitly distribute the application. Here, we might utilize the existing user base of the social VR platform, or attempt to attract new users to it.

\subsubsection{App Stores}

Application stores offer another, potentially effective way to distribute VR applications. Here, applications are published in app stores, such as Google Play\footnote{\url{https://support.google.com/googleplay/android-developer/answer/6334282?hl=en}, last accessed January 21, 2021},
Steam\footnote{\url{https://partner.steamgames.com/doc/store/creator_homepage}, last accessed January 21, 2021}, or the Oculus Go Store\footnote{\url{https://developer.oculus.com/distribute/latest/tasks/publish-submit-app-review/}, last accessed January 21, 2021}. This approach poses certain requirements to researchers. Firstly, app stores might require the application to be developed using certain tools and programming languages. Secondly, applications are usually subject to a review and, hence, researchers must fulfill certain criteria regarding data protection, user interface design, etc. Thirdly, publication on app stores are often not free of charge. 

At the same time, publishing in app stores is attractive because a large audience can be reached. Indeed, prior research has successfully distributed research prototypes through app stores in more conventional contexts like mobile interaction \cite{Henze10,henze2011,Henze11niels}. Furthermore, since the application stores take care of the installation process, it tends to be safe, free of errors, and easy for consumers.

We provide a brief overview of some popular application stores and their viability for publishing applications for research purposes. Our list is not intended to be exhaustive, but rather, to provide a general idea of the possibilities and challenges with app stores. Other app stores might be available or emerge in the future for research purposes.

\begin{description}
\item[Steam] is an app store particularly popular with games and VR games that support SteamVR. One caveat of Steam is its strong focus on (commercial) games. There are only a few exceptions, such as software for content creation. Nevertheless, it might be interesting if the research prototype can be embedded into a game, which is particularly popular in the HCI community. A strong advantage of Steam is that it has millions of users and that many HMDs supporting SteamVR exist, hence allowing a large audience to be reached. Researchers need to be aware that publishing takes at least 30 days and has a charge of about 100\$ per application. 

\item[The Oculus Store] is targeted at Oculus devices and, hence, cannot be used by owners of some other popular HMDs. At the same time, the consumer base is fairly large. The store provides clear guidelines for content publishing and the review process takes around two weeks. After review, the content may be published live on the Oculus store or it can be shared via a download key with participants. 

\item[The Google Play Store] is immensely popular, connecting developers to millions of Android users. The Play Store allows for publishing Android-based VR apps, which can be run on a few HMDs and, more importantly, on any Android smart phone serving as a low-cost HMD in a cardboard. A Google Play developer account is needed, which is available for a one time fee of 25\$. One can choose to publish based on several available modes (internal test, closed test, open beta, full publish), called tracks. There are also other options to control the availability of the app, like country-specific availability. These options are very useful for researchers who might want to retain some control over how many people install their app or who they recruit as participants. Overall, the Google Play Store is a highly viable option for researchers, with the downside of not being an option for most high-end VR HMDs.
\end{description}

\subsection{Summary}

We identified several different ways to conduct remote VR studies. On one end of the spectrum, researchers can produce isolated VR prototypes and distribute them directly to participants. On the other end, researchers can set up their studies in existing VR platforms and make full use of their features. At the same time, there are options in-between, like distributing a custom-built application through app stores, or uploading it to a VR platform that supports custom applications.

These approaches offer a good starting point for our investigation. It is worth noting that there are other considerations involved as well, such as how participants can best be recruited for remote studies. We will get back to these considerations in the later chapters, when we synthesize our findings from the online survey and the two case studies.

With these possibilities in mind, we then conducted the two case studies, which we present next. For these studies, we identified the best ways to run them remotely, which were very different from each other. This also allowed us to report on a wide range of experiences.

\section{ Case Studies}

To complement the online survey as well as our investigation of different approaches, we present two case studies to gather first-hand experience. The studies we present are part of independent research projects, in which some of the co-authors of this submission are involved. In this paper, we focus on aspects related to the \textit{study methodology} in the description of the case studies, rather than on the main research questions of the projects and the obtained results.


\subsection{Case Study 1: Health Interfaces in VR Shooters}

\begin{figure}
    \centering
    \includegraphics[width=1.0\textwidth]{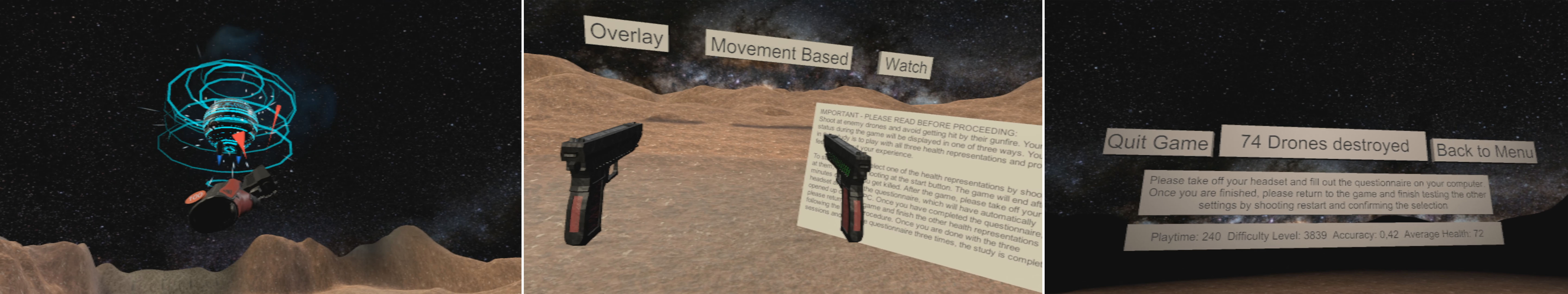}
    \vspace{-5mm}
    \caption{The VR shooting game and instructions used in Case Study 1. \textit{LEFT:} players shoot drones using their VR contollers as pistols. One of the tested health interfaces, the watch, is visible on the user's wrist. \textit{MIDDLE:} The starting instructions and a menu where the players choose which health interface to use. After their selection, they can begin the game. \textit{RIGHT:} Instructions and menu displayed after playing a session with one of the health interfaces. Study participants repeated this procedure three times.}
        \vspace{-4mm}
    \label{fig:health_interfaces_vr_shooters}
\end{figure}

In Case Study 1, we developed a VR shooting game (Figure \ref{fig:health_interfaces_vr_shooters}) where study participants attempted to shoot at drones using their VR controllers, and avoid getting hit by the drones' lasers. Our research focus was on investigating different ways to convey the player's \textit{health}. The player's health was reduced each time they got hit by lasers, and would slowly restore if they did not get hit for enough time. The game ended when health was reduced to zero.

We designed and implemented three different interfaces to communicate the player's health status. Study participants played three rounds of the game, each time with a different health interface, and provided feedback.

In this study, the most critical aspects of the remote VR methodology were:
\begin{itemize}
    \item \textbf{Standalone VR application.} We built a custom, standalone VR application. We did not use existing platforms like Rec Room or VRChat.
    \item \textbf{Direct download.} We hosted the application on a web server and provided a link to it to participants. Participants then downloaded the application and ran it on their own.
    \item \textbf{Independent study method.} The study was asynchronous, i.e., participants ran the study on their own any time they wanted. No experimenter was present during the study.
\end{itemize}

\subsubsection{Study Description}

We investigated whether diegetic representations of player's health status in VR shooting games would improve the sense of presence and add to a sense of danger during intense gaming sessions. As a non-diegetic baseline condition, the player's health was displayed as a health bar, which was visible on the screen at all times. As diegetic conditions, we implemented a wristwatch displaying the player's health (players would need to lift their wrist and really look at the watch) (Figure \ref{fig:health_interfaces_vr_shooters}, left) and a movement-based method, where hurt players moved slower and their firearm was shaking, to simulate the condition of being physically hurt. We hypothesized that these diegetic conditions would improve certain aspects of the gaming experience.

\paragraph{Study Design.} We built a custom VR shooter game using Unity and SteamVR. In the game's main menu, instructions were displayed for the player. They could choose any of the three health interfaces and start the game (Figure \ref{fig:health_interfaces_vr_shooters}, middle). In the game, drones appeared around the player which attempted to shoot at the player. The player's task was to avoid getting hit and try to shoot as many drones as possible. The game session was set to last for four minutes. To support players of different skill levels, we implemented a dynamic difficulty system that adjusted various factors based on how well the player performed: the rate at which the drones appeared, how spread out they would be, and how fast and frequently they would shoot at the player. Also, we added a curve to player health -- a typical trick in videogames -- so that players would not die as easily as the situation might suggest. These were important factors in ensuring that players of all skill levels would experience a challenge and the thrill of being low on health, which in turn was important for our research goals. At the same time, the dynamic difficulty and the curved health would not let players die very easily and frustrate them. Players regained their health during the session if they remained unhurt for a brief period.

Participants played a short session using all three health interfaces. After each session, a feedback questionnaire was automatically opened on their PC on their default browser, and the game instructed participants to take off their HMD, fill in the form, and then return to the game (Figure \ref{fig:health_interfaces_vr_shooters}, right). We also used the form as a means to log data from the game: we pre-filled some of the form questions with game data and prevented users from modifying them.

Even though current research suggests filling questionnaires inside VR \cite{Alexandrovsky2020QuestionnairesVR,schwind2019using}, we opted for external questionnaires. This was partly because we wanted to give players a proper break after a potentially intense game session, but more importantly because the questionnaires were so long that they would have been very tedious to fill in VR. The questionnaires after each session included the Game Experience Questionnaire \cite{ijsselsteijn2013GEQ}, the Slater-Usoh-Steed presence questionnaire (e.g.,  \cite{slater1998influence,Usoh1999WalkingVE}), custom statements, and also some optional open-ended questions.

When participants filled in the form for the third and final time, they ticked a checkbox stating that they had finished playing with all three health interfaces. At this point participants were presented with a brief set of additional questions about their background and their experience about participating in a remote VR study.

We took special care to provide clear instructions to participants and support them throughout the study process. In our advertisement, we provided a brief overview of what the study would contain. In our recruitment email, we had instructions for downloading and running the VR application as well as the high-level steps they should take in the study. Inside the VR game, we had a text box containing the same instructions, which the users would see immediately after starting the application (Figure \ref{fig:health_interfaces_vr_shooters}, middle). Further, there was a brief explanation about each health interface before they started the game. After the game, there were instructions to take off the HMD and fill in the questionnaire on their computer, and then return to the game (Figure \ref{fig:health_interfaces_vr_shooters}, right). The same reminder was added at the end of the questionnaire, that told participants to return to the game.

In addition, we set up a separate issue report form, that prospective participants were encouraged to submit in case they experienced any problems with the game or the study. This issue report form was added in the advertisements as well as in the emails that contained the download link.

\paragraph{Recruitment.} We did two rounds of recruitment. In the first round, participation was voluntary, i.e., no rewards were offered. We advertised the study through channels we had identified as suitable for this research, such as Reddit. We also advertised the study to a practical course about VR technologies and programming that took place at the same time at a local university. Some of the students owned a private HMD, and some had been loaned an HMD by the university.

In the recruitment call, we included instructions for the study procedure (even though all the instructions were in the game as well), because the study was designed to be run fully independently and no experimenters were present at any point. It was therefore important for participants to know exactly what to expect in the study. The call also included a direct download link to the game as well as a link to the issue report form.

In the second round, we advertised the study through the same channels with a 10€/12\$ reward, paid via PayPal. We removed the download link from recruitment calls and instead set up a separate registration form where participants provided basic user information as well as an email address. We then sent the study instructions along with the download link to the registered participants. We did this to retain control, so that no one would try to cheat (e.g, by completing the study several times), and that we would not be flooded with too many participants who all needed to be compensated.

This way, we recruited 24 participants (18 male, 5 female, 1 undisclosed). Their average age was 24 (SD = 7.9). The participants used a variety of different VR devices: HTC Vive Pro (7), Valve Index (7), HTC Vive (6), Oculus Quest (2), Oculus Rift S (1), and Asus Windows Mixed Reality Headset (1).

The participants also reported their estimates on how much they play VR games on an average week, and how much they play other digital games. With regards the VR gaming, 12 participants (50\%) reported that they play 0--5 hours on an average week. Seven (29\%) estimated 5--10 hours, four (17\%) estimated 10--15 hours, and one participant (4\%) estimated that they play 20+ hours per week. With regards to other digital games, seven participants (29\%) estimated that they play 0--5 hours a week, eight (33\%) estimated 5--10 hours, three (13\%) estimated 10-15 hours, two (8\%) estimated 15-20 hours, and four (17\%) estimated that they play 20+ hours per week.

\subsubsection{Participant Experience}

In addition to collecting data to answer our study-specific research questions, we gathered impressions on how the participants felt about the remote study. We asked them to respond to three statements on a 7-point Likert scale, where 1 = strongly disagree, 4 = neither agree nor disagree, and 7 = strongly agree.

Participants were very positive about their experience, strongly agreeing with all three statements (MD = 7). They felt that \textbf{(1)} they were comfortable participating from home, \textbf{(2)} the instructions for how to run the study independently were clear, and \textbf{(3)} they did not need anyone in the same room to consult. One participant wrote a comment, commending us for a "smooth remote study".

The positive feedback from participants is encouraging. It supports our findings from the online survey, where participants felt comfortable about the thought of remote participation. We were somewhat surprised by this. Rather, we expected that the fully independent study procedure might make some participants stressed or uncomfortable, but this did not seem to be the case. This strengthens our belief that our procedure was well thought out and our instructions prior to and during the study were clear (as also reported by the participants). Still, it is possible that the fully independent procedure drove away some potential participants in the recruitment phase.

In addition to the clarity of instructions, we believe that convenience and familiarity played key roles in the positive experience. In remote studies, users do not have to spend money and time traveling to a lab and they do not have to get familiar with new equipment or meet with new people. In contrast, users can participate from home. This familiar setting likely provides support and safety. They also use their own equipment that they are familiar with before the study.

\subsubsection{Experiences and Discussion of Running a Remote Study.} In the first recruitment round, we faced a major challenge with getting a sufficient number of participants. Logically, this was because we did not compensate participants, and we required a SteamVR-compatible HMD, so low-end devices were not suitable. A total of 10 people participated in the study. Five of them came from the practical university course, three came from Reddit, and two from other channels.


In the second round, we immediately received 33 registrations overnight, which resulted in us having to close the recruitment form. All who registered were sent the instructions, and 15 of them participated. Therefore, offering a reward solved the recruitment issues. Had we kept the registration form open for longer, we would have likely been able to recruit many more participants, despite the pre-requisite of a SteamVR-compatible HMD. A separate registration form also seemed to be an good choice for controlling the number of participants. Direct access to the study could result in issues for researchers, who often have a limited budget for rewarding study participants.


One of our initial concerns was that we would suffer from a high drop-out rate (participants quitting mid-study). However, out of the total 25 people who ran our VR application, 20 participants (80\%) completed the study in full. Four participants (16\%) completed two of the three conditions, and one participant completed only one condition. We see this as a very high completion rate for an entirely remote and independent study procedure that took approximately 30--40 minutes. We believe that this success was due to our efforts to provide as clear instructions as possible for the participants to perform independently, and to make the transitions between study phases as effortless as possible (e.g., moving from the game to a pre-opened, pre-filled questionnaire).

However, the number of second-round participants (15, of which 12 completed the study in full) was low compared to the number of registered prospective participants (33). It is difficult to say why so many registrants did not follow up in the end, but our experience would indicate that researchers should prepare for this happening.

Three people reported issues through our issue report form. Those were of technical nature. Two participants, who both had Valve Index controllers, reported that they were not able to shoot using the triggers, but they were able to shoot by pressing other buttons. One participant reported that their pistols were slightly misaligned with their hands. The reported issues were minor, as participants were still able to complete the study. Still, such instances are testament to the challenges with remote studies, as things that researchers typically have control over (e.g., calibration, testing, solving issues on the spot) are outside of their grasp. 

\paragraph{Conclusion.} Besides initial recruitment issues (which we overcame later), the study procedure worked very well. We did not identify major pitfalls in our procedure, which we had designed carefully, keeping in mind that the participants do not have anyone to consult during the experiment, and that they must be motivated to finish on their own. Furthermore, the participants gave very positive feedback about the remote study.

As we could expect based on our survey data, our participants were dominantly male (75\%). Although there is a prevalent gender imbalance in HCI studies in general \cite{Caine:2016:LocalStandardsSampleSize}, this bias is still noticeably strong. The average age of our participants was 24 (SD = 7.9). This is indeed not high, but it is likely no lower than what HCI studies typically have. While we are not aware of an average participant age across HCI studies, we do know that study participants are typically students \cite{Caine:2016:LocalStandardsSampleSize}, and they likely fall in this range. It is also worth noting that our standard deviation was relatively high (7.9), with some participants being over 40 years old.
With respect to the participants' gaming habits, most were rather moderate with their VR use (0--5 hours or 5--10 hours per week). Instead, they were more actively gaming on other devices (e.g., PC, consoles). This is not surprising, considering the potential fatigue in VR gaming as well as the fact that VR gaming is not yet as mature as other forms of digital gaming. In any case, our participants seemed to represent a relatively balanced group of moderate games and hardcore gamers (judging from their play time).

\subsection{Case Study 2: Investigating Proxemics in Virtual Reality using Rec Room}

\begin{figure}
    \centering   \includegraphics[width=\textwidth]{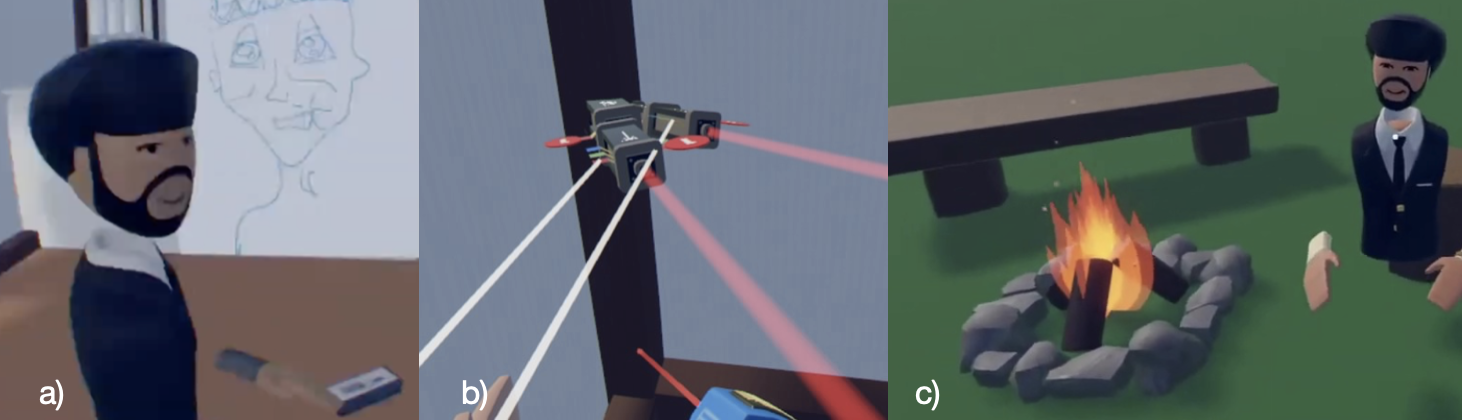}
    \caption{Proxemics study (Case Study 2). {a:} Players playing inside the VR game. \textit{b:} Range finder feature of Rec Room (used to calculate the distance between players). \textit{c:} Conducting interviews at the end of the study.}
 \label{recstudy}
\end{figure}

In Case Study 2, we set up a VR environment (Figure \ref{recstudy}) where two participants completed collaborative tasks. Our research focus was on investigating how the gender of virtual avatars affects the interpersonal distance (IPD) between people. We measured the distance between the participants at specific points during the study.

We designed a virtual environment in Rec Room and implemented two different collaborative tasks, as well as implemented a sophisticated system to track the distance between participants. Participants switched their avatar genders halfway through the study.

In this study, the most critical aspects of the remote VR methodology were:
\begin{itemize}
    \item \textbf{VR application built within an existing VR platform.} We built a custom VR environment in Rec Room, using the tools provided by the platform.
    \item \textbf{Distribution through the VR platform.} Because our application was built inside Rec Room, there was no need to download or install an application. Instead, our study was accessible through Rec Room.
    \item \textbf{Remotely guided study method.} A remote experimenter was present during the study to recruit users  and to set up and manage the study environment in each session.
\end{itemize}

\subsubsection{Study Description}

We investigated proxemics in virtual reality using Rec Room. The proxemics theory studies the interpersonal distance between humans during interaction \cite{DANESI2006241}. We explored to what extent this theory applies to virtual reality. In particular, we investigated whether the relationship between people (strangers and friends) as well as gender  affects proxemics.

\paragraph{Study Design.} 
We designed a collaborative two-player game in which participants had to solve puzzles together to unlock rewards (Figure \ref{recstudy}a). To understand the influence of the relationship between people on proxemics, we recruited both pairs of friends as well as pairs of strangers. To additionally investigate the role of gender, the study consisted of two phases. In the first phase, participants collaborated on the puzzle-solving task with an avatar matching their gender. In the second phase, participants were represented by an avatar of the opposite gender. The study consisted of two sessions. After the first session participants were required to go back to their own room and change their avatar gender and then return to join for the second session of the study.

The experimenter was present in the room throughout the study. The experimenter embodied their own avatar in the study space, provided instructions to participants and oversaw the procedure, very much like an experimenter would do during a lab study.

The distance between the players was logged using the Rec Room feature "Range Finder" (Figure \ref{recstudy}b), that can track the distance from a certain point to a player. However, because Rec Room does not allow storing such information, we had to implement a workaround. We set up the Range Finder data on a virtual display in Rec Room and recorded the data using screen capture. We then used Optical Character Recognition (OCR) to extract the data.

After the players finished the game, they filled in a demographics questionnaire and the experimenter interviewed them inside Rec Room (Figure \ref{recstudy}c). Different rooms were designed for each purpose so participants played the game in one room and completed the interview in a separate room. Participants were compensated with 5€ via PayPal.


\paragraph{Recruitment.}

We distributed a call for participation via various social media platforms, university mailing lists, in the Rec Room community on Reddit, and via Facebook. In addition, we reached out to people active in the Rec Room community. In particular, one moderator from the Reddit Rec Room group advertised the study via social media (Facebook). One Rec Room tutor, who hosts a tutorial class on circuit and logic design inside the platform, suggested this study to his followers.  Furthermore, we asked participants to advertise the study to friends who might be interested. This led to additional participants. Finally, we recruited active players in Rec Room on the fly. This turned out to be an effective method, since those players were already in Rec Room and were often available for participation.

For this study, we struggled to find female participants. This led us to have an imbalanced gender ratio and, therefore, we could not conduct studies between three gender groups (male-male, male-female, female-female) as planned initially. Eventually we had only male-male pairs (15) and male-female pairs (5). Therefore, we had 40 participants (35 males) with an average age of 23.7 years (SD = 7.75). In general, we found the recruitment process to be fairly easy for the Rec Room study. One drawback was a difference in time zones between the experimenter and the participants, and thus the experimenter had to conduct some sessions at unusual hours (e.g., very late at night).

\subsubsection{Participant Experience}

In addition to collecting data to answer our study-specific research questions, we gathered impressions on how the participants felt about the remote study. Hence, in the interview at the end of the study, we asked them open-ended questions about their experience.

All 40 participants were overall very positive about the experience. All of them stated that they felt comfortable about participation in a remote study. Furthermore, 39 participants stated that the remote study in Rec Room was a novel experience to them; one participant mentioned to have previously attended a focus group in Rec Room. We asked them if they faced any challenges in understanding instructions or in participating. For all participants, it was an easy experiment and they did not face any difficulty during the experiment. 


Similar to case study 1, it is likely that participants felt positive about the experience due to the convenience (participate from home, no need for travel or other inconveniences) and familiarity (familiar and safe home setting, familiar equipment and platform). The difference to the first study was that we used Rec Room with a remote experimenter, which also seemed to function very well from the participants' perspective. It is notable that Rec Room is a social platform where participants likely had many interactions with other people before. Therefore, interacting with a stranger (the remote experimenter as well as the other participants in the stranger condition) was not new to them.

\subsubsection{Experiences and Discussion of Running a Remote Study}

In total we recruited 40 participants in pairs. As many friends play together in Rec Room it was easy to recruit friends in pairs. It was also easy to recruit strangers in pairs. For recruitment, the experimenter would generally join playing rooms and approach other players about the study. Since each game room consists of many players, both friends and also strangers playing altogether, recruiting was easily possible. 


Our main challenge was the balanced recruitment of male and female participants. Ultimately, we failed to do this, as recruiting female-female pairs proved to be difficult. Even though we otherwise ran a very successful remote study, the lack of female participants limited our findings regarding gender differences. While we also faced a bias towards males in the first case study, the issue was more clear in this second study. This is likely due to two factors. First, we needed participants to attend the study in pairs; finding two female participants for one session as opposed to just one female participant is naturally more difficult. Second, according to your survey the general pool of HMD owners (and likely Rec Room users) is male-dominated.

One important aspect to consider is that the Rec Room community consists of a large number of users below the age of 18. Hence, proper verification of the participants' age should be put in place upon recruitment. At least, participants should be told that they are required to be at least 18 years of age to be eligible to participate. But this also makes it easy for researchers to recruit underage participants if required for the study. In this case, researchers would need to find ways of obtaining consent from participants' parents. 

One challenge during recruitment and conducting the study sessions was that potential participants were spread across the globe. Therefore, some participants were not available at regular working hours for the experimenter. As a result, the experimenter ran some sessions in the middle to accommodate for different time zones.

We planned for position tracking in VR to measure the interpersonal distance of participants. Rec Room allows the position of each player to be visualized; however, logging this data was not possible. We worked around this by recording the visualization during the study and then using optical character recognition to extract the numbers. This worked well in the end, but required some experimenting to get right, so that OCR would produce the correct output. Depending on the study being conducted, implementation challenges are likely to occur when using existing VR platforms, as their features can be limited. 

Another design challenge specific to Rec Room was the limited ink\footnote{\url{https://rec-room.fandom.com/wiki/Maker_Pen}} available for designers. The pen system in Rec Room allows creators to design a room and this has an ink limit to facilitate robust processing, i.e., the rooms have a limit to how many objects they can contain. Thus, when designing the room, we had to plan economically. 

The study was interrupted a few times when the participant's game crashed and they left the room. Additionally, since the experimenter was also virtually present wearing an HMD device, it was difficult for the experimenter to check their personal notes during the study.

\subsubsection{Conclusion}

After some creative workarounds, our overall study procedure worked very well. We had no problems during the study sessions, the participants' were positive about their experience, and our data collection methods worked as intended. Recruiting participants turned out to be easy in several ways. Most importantly, the option to directly recruit participants from inside Rec Room was very helpful.

As noted earlier, though, it was critical for our research to also recruit female pairs as participants, and this was a major challenge. This showcases the limitations in the demographic of HMD users, although in most studies the issue should not be as pronounced as here. Two minor inconveniences were that some participants were in far-away time zones, and that Rec Room had limited data collection features, so we could not record the exact distance between participants directly. Researchers should be prepared for creative workarounds in such cases.

\section{A Framework for Running Remote VR Studies}

\begin{figure*}
  \includegraphics[width=\textwidth]{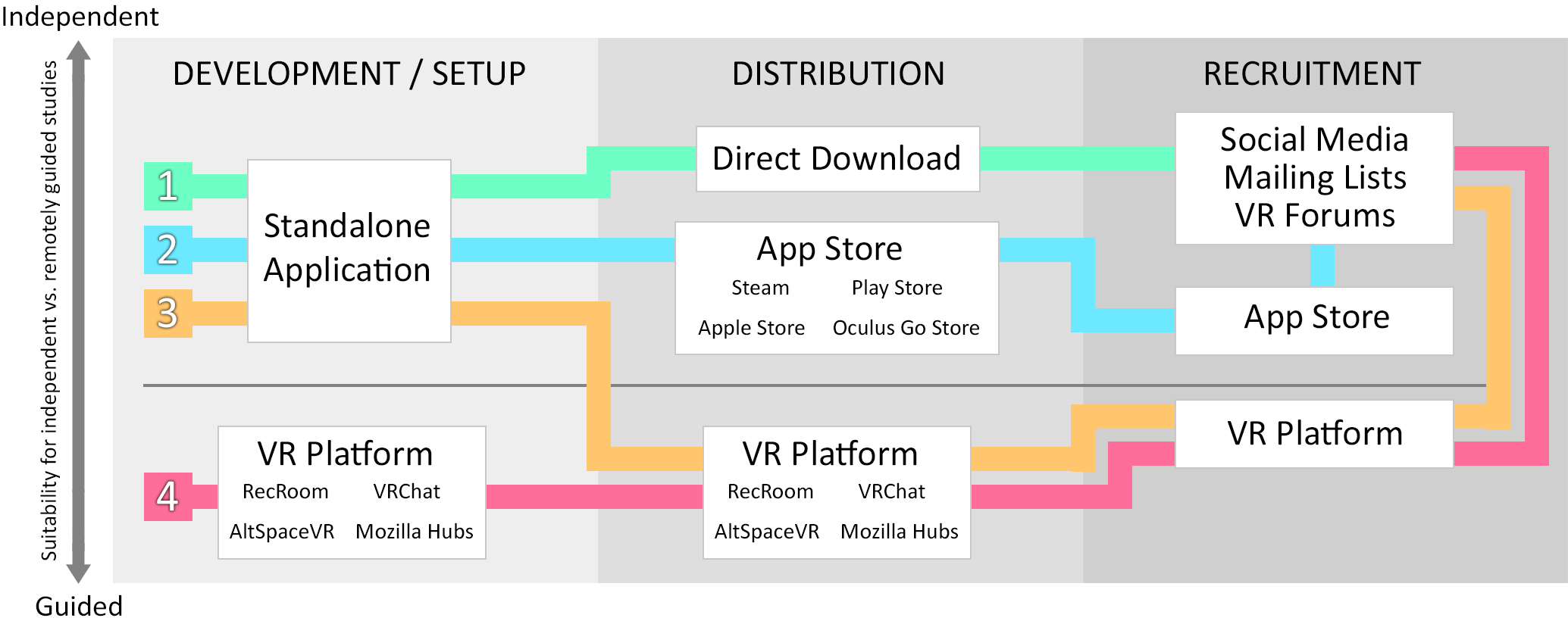}
  \caption[width=\linewidth]{We envision four high-level procedures to conducting remote VR studies. Paths 1 and 2 are more ideal for studies that participants complete independently, while Paths 3 and 4 are better suited for remotely guided studies. \textit{PATH 1:} Standalone VR application that participants download and install independently. \textit{PATH 2:} Standalone VR application that is published in an app store. \textit{PATH 3:} VR application that is uploaded to an existing VR platform, which can also be used to recruit participants. \textit{PATH 4:} VR application that is built using the built-in tools of an existing VR platform, which can also be used to recruit participants.}
  \label{fig_remote_vr_study_framework}
\end{figure*}

In this section, we synthesize the findings from our online survey, the analysis of different approaches (standalone, app store, VR platform), and two case studies into a framework for remote VR studies (Figure \ref{fig_remote_vr_study_framework}). We discuss the strengths and weaknesses of each approach, and when each approach might be appropriate. The framework is meant to assist researchers in planning their own remote VR studies.


\subsection{Four Primary Approaches to Remote VR Studies}

We distinguish four primary approaches to conducting remote VR studies -- we refer to these as Paths 1--4. In the following, we present these four approaches and discuss their differences. In particular, we discuss whether each path is more suitable for independent studies that participants run without an experimenter, or remotely guided studies, where an experimenter is present. We also discuss the strengths and weaknesses of these approaches.

\subsubsection{Path 1: Standalone Application + Direct Download}

In Path 1, which we experimented with in case study 1, researchers build a standalone, independent VR application, which participants download and install themselves. The study can then be advertised through various channels. In our experience, VR-related forums (like VR subreddits in Reddit) worked best in attracting HMD owners.

The strengths of this approach are that independent applications can have the most extensive functionality and the most options for collecting data, because the applications do not need to conform to the limitations and regulations of app stores and VR platforms. In our study, we logged extensive quantitative and qualitative data with relative ease.

Still, this approach might not be ideal for every development platform. For example, installing external applications for Android devices requires extra steps, such as turning on developer mode, which not every consumer is willing or even capable of doing. In our case study, this worked well, as participants only needed to download our game package, after which they could immediately run the game through an executable.

Researchers should consider whether or not they should distribute the download link openly. In case participants are rewarded, researchers might consider mechanisms to control that not too many participants take the study (e.g., so that their budget is not exceeded), and to reduce the chances of people "hacking" the system (e.g., by completing the study and claiming the reward multiple times). Researchers could, for example, set up a website for pre-registration, and only registered participants would receive the download link. In our case study, this approach worked well, although it is likely that not all pre-registrants will complete the study.

We believe Path 1 to be particularly suitable for independent studies, where participants go through the study procedure alone, without consulting an experimenter. In our case study, this approach worked extremely well, although the procedure should be planned meticulously. Extra care should be paid to providing clear instructions. In best case scenarios using Path 1, the VR application is distributed and advertised, after which the entire study runs on its own.

It is certainly possible to run experiments with a remote experimenter through Path 1, but this requires additional steps like agreeing on specific time slots with each participant. Depending on the study, a remote connection to the VR application would need to be developed for the experimenter, or a connection would need to be established via other means like video conferencing tools (which would, in turn, limit how much control the experimenter has and what insights they can gather).

\subsubsection{Path 2: Standalone Application + Publishing through App Store}

In Path 2, researchers build a VR application and publish it in an app store. In some cases, it is possible that the app store itself attracts participants, but researchers should be prepared to reach out to potential participants through additional channels.

The main strengths of this approach are that distributing and installing the application is easier and less error-prone (as opposed to independent installation), and might help in reaching a wide and diverse audience. App stores have been commonly used in other research areas, for example, in mobile interaction \cite{Henze10,henze2011,Henze11niels}, where large participant numbers were reached. Opportunities for data collection are still relatively extensive, although somewhat more limited than in Path 1, as the application must conform to the rules and regulations of the app store.

Based on our analysis of existing app stores as well as own experience with publishing a VR app on Google's Play Store, there are considerable differences between app stores in terms of how easy it is to publish apps (some require a review by the provider) as well as how much effort and time they require (for some of them it takes up to 1 month to receive feedback). Researchers should inform themselves prior to choosing a particular app store. 

For the same reasons as Path 1, we believe that Path 2 is more suitable for independent studies; in the best case scenario, studies using Path 2 could also run on their own after being advertised.

\subsubsection{Path 3: Standalone Application + Upload to VR Platform}

In Path 3, researchers build a VR application and upload it to an existing VR platform, such as VRChat. The study can be advertised through the platform itself, but also through conventional channels.

The main strengths of this approach come from the social aspects of the VR platforms. They are built for multiple users to connect to, and interact with, each other. As such, many things critical to some users studies are already there or can be implemented with little effort. For example, experimenters and additional users can connect to and interact with the environment without additional effort. In addition, the VR platforms serve as excellent recruitment channels, evidenced by our experience with Rec Room and a prior study in VRChat \cite{Saffo:2020:CrowdsourcingVR}. Experimenters can look for participants within the platform and recruit them on the spot. These strengths make Path 3 well-suited for remotely guided studies where an experimenter is present, and perhaps particularly suitable for studies where multiple participants are needed simultaneously. 


Publishing custom applications in VR platforms requires the application to follow certain regulations, and requires that the application is implemented correctly (e.g., that it uses specific APIs and libraries). This somewhat limits the features that can be included in the application and what kinds of data can be collected. On the other hand, connecting an experimenter allows them to easily conduct in-depth observations and interviews, which are more difficult with Paths 1 and 2. Using a VR platform also helps in overcoming compatibility issues.


\subsubsection{Path 4: Setup directly in VR Platform}

In Path 4, researchers build the VR "application" directly within an existing VR platform, such as Rec Room, using the platform's built-in tools (like we did in case study 2). The study can be advertised through the platform itself, but also through conventional channels.

Similar to Path 3, the strengths of this approach are that experimenters can easily connect to the study sessions and the platforms serve as excellent recruitment channels. A unique strength of Path 4 is that the tools offered by the platforms make setting up study environments fast and easy, as opposed to implementing custom VR apps. For example, Rec Room offers a simple user interface for building rooms inside the platform, adding and modifying objects, and setting up interactions.

Still, the possibilities offered by VR platforms are limited, as they are rarely built for research purposes, but for simple social interactions between users. Therefore, when considering which path to take, researchers should look into existing VR platforms to see whether any of them offer the required functionality. A related limitation is that VR platforms offer limited ways to collect (especially quantitative) data from users. Still, it is often possible to build questionnaires within the platforms (like we did in Rec Room). Moreover, like in Path 3, connecting an experimenter allows in-depth observations and interviews to be conducted.

\subsection{Choosing the Best Approach for a Remote Study}

The four approaches balance different needs for user studies. While most approaches can be adapted to fit different studies one way or another, there are still general considerations that can be used to identify the best approach. It is also worth noting that for certain VR studies, remote approaches might not be feasible at all, or they might present significant challenges. Below, we discuss some perspectives for evaluating which remote approach is the most feasible:

\subsubsection{Consider Your Requirements for Interaction}

In deciding which approach to take with a remote VR study, researchers can consider the interactive features that their study requires, and investigate whether existing VR platforms meet their criteria. We believe that Path 4 is generally the easiest path to take, and, therefore, it might be useful to evaluate whether Path 4 is a feasible option in the first place. Path 1 offers the most control over what interactive features can be included, with Paths 2 and 3 situating in the middle, being also dependent on the targeted app store or VR platform.

\subsubsection{Consider Your Requirements for Data Collection}

Similar to the required interactive features, researchers should consider the data that they need to collect. Remote studies in general are more limited than lab studies, and collecting data through specialized hardware such as biosensors, electroencephalography (EEG) and electromyography (EMG) is very challenging. Despite current technology making a greater range of data available (such as eye tracking data), additional hardware requirements constrain remote studies. Hence, remote studies are not always possible.

Setting specialized hardware requirements aside, more typical data can certainly be collected remotely. Much like with with interactive features, Path 1 offers the most control over what data can be logged from the application. Path 2 also offers much control over logged data, although some app stores might impose some limitations. Paths 1 and 2 are therefore well suited particularly for gathering quantitative data.

Because VR platforms limit data logging -- for privacy and safety reasons -- Paths 3 and 4 are less ideal for collecting quantitative data. However, some limited options may still be available, and also some workarounds exist. For example, as demonstrated in case study 2, we were unable to log participant positions from Rec Room. We obtained the required data using screen capture and OCR to extract the data. This was a successful workaround, yet required additional effort from us in terms and planning and experimentation, and we also had to purchase additional OCR software. 

In contrast, with qualitative data the advantages may be flipped. Because Paths 3 and 4 are well suited for studies where an experimenter is present, they are --- by extension --- suitable for collecting qualitative data through observations and interviews. Collecting qualitative data through Paths 1 and 2 is certainly possible, but may require additional effort. For example, for standalone applications, the possibility for a remote experimenter to connect to the same session as the participant must be implemented, or they must use additional software to connect for the sessions. Video conferencing tools may be helpful here. However, despite their ease of use they still add additional steps to the procedure and might not be enough to capture all details in the session.

Therefore, studies relying heavily on quantitative data (or the collection of large amounts of data) might benefit more from Paths 1 and 2. For studies with a qualitative focus, Paths 3 and 4 could be considered. We again emphasize that these are not set in stone. For example, with Paths 1 and 2, video conferencing tools, screen sharing tools, etc. could be used to connect with participants, but they are still limited and require extra steps in the procedure.


\subsubsection{Consider the Advantages and Drawbacks of Independent and Remotely Guided Studies}

As already discussed, independent studies likely support quantitative data, while remotely guided studies might support qualitative data. There are also other considerations to these two approaches.

For independent (asynchronous) studies, the benefits are that since the actual procedure requires no resources, the study can run day and night and potentially reach a large number of participants. For certain studies, it is also considerably easier to develop a VR application if one does not have to worry about functionality surrounding the experimenter. However, due to the lack of an experimenter, independent studies might be more prone to malfunctions (e.g., bad calibration, hardware issues) and misunderstandings (e.g., not following the procedure correctly), or participants might simply be overwhelmed or discouraged by the procedure.

In contrast, through a remote connection, experimenters can instruct participants and oversee that the procedure is followed accordingly. Especially, they can tackle any unforeseen events and provide support if problems occur, and conduct observations and interviews. Remote guidance might be particularly desirable if the participants need to learn or train new things, if the procedure has several phases with instructions, or if the study has multiple things running in parallel, like several participants. A case in point is our case study 2, where participants attended the sessions in pairs. This study would have been very difficult to run without an experimenter who was required to monitor each session and instruct participants accordingly, especially when considering the limitations of the used VR platform. The obvious drawback is that an experimenter is then tied to running each study session, which takes time. Experimenters might also face challenges with different time zones; flexibility with schedules might be needed to reach participants located in other countries.



\subsubsection{Reflections on the Longevity of the Framework}

Here, we briefly reflect on how we believe this framework and the different approaches to remote VR studies might evolve in the long term. It is inevitable that the VR landscape, the tools, and the user base will evolve. However, we have formulated most of our findings and considerations in such a way that they will still hold despite the changes that we expect to see in the future.

We have only recently arrived at a situation where a notable enough number of people own head-mounted displays. But VR technologies and the related tools are still relatively immature, and the number of people who own HMDs is still marginal to that of mainstream devices. We believe that in the future, remote VR studies in general will become easier, as VR technologies develop and a larger audience becomes equipped and acquainted with head-mounted displays---and possibly other VR devices in the far future.

Through the future advancement of VR technologies, we may eventually have more data collection opportunities in remote studies. However, we believe that studies using app stores and especially studies using social VR platforms will continue to be limited in their data collection. Such limitations are in place for good reasons---to protect the privacy and safety of the users---so this is unlikely to change. Hence, Path 1 (standalone application + direct download) might benefit the most from advanced data collection, which we believe is already the most flexible in this regard.

Data collection limitations aside, it is reasonable to assume that social VR platforms will continue to develop alongside HMDs, and offer more advanced and comprehensive interactive options as well as tools for content creators. Therefore, we believe that Paths 3 and 4 (that utilize social VR platforms) will become a feasible option for a wider range of studies.



\section{Best Practices \& Lessons Learned}

In this section, we draw upon our results and experiences, and provide some best practices and lessons learned that we hope help researchers while planning their remote VR studies.


\subsection{Understand HMD Users and Demographic Limitations}

Even though it is possible to reach a large number of participants remotely compared to a lab study, remote studies may suffer from certain demographic limitations. As we learned from our survey, HMD owners are biased towards young males. This is also evidenced by our two case studies, where the majority of participants were male (75\% and 87.5\%, respectively).

The bias towards male participants was particularly a challenge in our second case study, where we initially attempted to investigate gender differences. However, due to the difficulties of finding female pairs, we had to exclude them and opt for an investigation between male-male pairs and mixed gender pairs. Hence, we received only limited insight into gender differences.

It is worth noting that the bias towards males is not necessarily solely due to the imbalance among HMD owners, but also due to the gender imbalance in suitable recruitment platforms (e.g., Reddit, Rec Room). In any case, researchers looking to utilize HMD owners as study participants should be aware of this limitation, particularly if gender balance is important with respect to the insights of the planned study.

As an advantage, however, online recruitment platforms allow pre-selecting participants based on demographics (e.g selecting participants who own a VR HMD). Recruitment in online VR platforms is easy as experimenters are likely to find available players online.


\subsection{Home VR Setups May be Small}

According to our survey, many home VR setups are smaller than what researchers might be used to in lab conditions. 29\% of respondents reported a small space of up to only 5\,$m^2$, while another 31\% reported a space between 5 and 10\,$m^2$. Therefore, it might be challenging to conduct studies that require more space. This may be especially true for user studies researching locomotion in VR or collaboration in social VR.  At the very least, researchers should be clear about their space requirements and implement mechanisms to ensure the safety of remote participants.

\subsection{Recruit Participants via VR Platforms}

In case study 2, where we set up our study in Rec Room, an effective way to recruit participants was to look for them from within Rec Room. Since Rec Room users are already using their HMD and are already within the platform, many of them were available for participation on the spot. This indicates that studies utilizing VR platforms (Paths 3 and 4), researchers might want to consider recruiting participants from within the platform.

\subsection{Identify the Rules of Subgroups in Forums and Social Media}

VR-related groups on various forums and social media platforms, especially Reddit, were also ideal locations for recruiting participants. However, many such platforms not only have their general community rules, but many subgroups have their own rules. Researchers should identify groups where posting study advertisements is allowed, and contact moderators in cases where the rules do not provide the necessary information.

\subsection{Set up a Channel for Participants to Report Issues in Independent Studies}

In case study 1, we set up a separate issue report form that we linked to in the original advertisement. Although we luckily did not run into any major issues, some participants still used it to report minor things that they had trouble with. Because of the lack of control and knowledge over how participants run the study, we recommend that researchers set up a similar channel for independent studies. This is particularly valuable with issues that altogether prevent a participant from completing the study, potentially making researchers aware of major issues early on.

\subsection{Ensure a Flexible Schedule in Remotely Guided Studies}

Potential study participants can be located anywhere in the world and hence they might be in a totally different time zone. In our case study 2, where we used a remote experimenter, this meant that some studies had to be run during the night to suit the participants' schedules. Of course, recruiting participants from several continents is not always necessary, but being able to recruit a more diverse set of participants is one of the strengths of remote studies.


\subsection{Make Notes Accessible for Remote Experimenters}

In our case study 2, an unforeseen challenge was that it was occasionally difficult for the experimenter to check their notes while wearing an HMD. There are several solutions to this. One practical consideration is for experimenters to have the notes \textit{on their person} instead of, e.g., on a table. In the latter case, experimenters may be disoriented when taking off the HMD, or they may be far away from the table, and may, therefore, need some time to locate their notes. A second consideration is that many VR platforms, such as Rec Room, can be used on desktop machines as well. This might be a more ideal way in some studies for the experimenters to connect from, instead of using an HMD. A third consideration is that it might be possible to put the experimenter's notes within the virtual world, so that the notes are only visible to them.

\subsection{Review Ethical Concerns and Ensure Transparency about Potential Risks}

VR technologies are associated with health and privacy risks \cite{spiegel2018ethics, bagheri2016virtual, Herz2019,Cobb1999}, for example, through cybersickness \cite{Cobb1999,Davis2014}, physical injuries \cite{menzies2016objective}), and effects on mental health \cite{aardema2010virtual,higuera2017}. The privacy risks of VR technology should also not be neglected, as large amounts of sensitive information may be involved, while access to and transmission of this data is often not transparently regulated \cite{DeGuzman2019,bagheri2016virtual}. These risks imply a need for reflections on the ethical consequences of VR studies, since they can affect participants. Remote VR studies could even aggravate the moral concerns as they are less controlled. 

Ethical concerns must be addressed to ensure safety and security of the participants. For example, the absence of an on-site researcher means no assistance to the participants in case of health issues. For remote studies, having a remote experimenter present during the study can improve the safety of the participants as the remote experimenter can monitor the participants and stop the study if required. Using internet to download study links or participating online may pose security threats to participants. We strongly suggest that researchers make their best effort to minimize such threats by using verified and trusted sources, such as official institution websites, to host study advertisements, information, and materials (like download links), instead of public sources such as links on Google Drive. Researchers can also focus on distributing studies in forums where safety measures exist (e.g., new accounts cannot make new posts, a certain trust level must be earned, or the community can help by marking safe and valid posts with badges or tags).

Therefore, we recommend a thoughtful consideration of all health and privacy risks and their mitigation. Moreover, transparent communication of risks to the participants is necessary to allow for an informed consent.

\section{Conclusion}

We investigated different ways to conduct VR studies remotely, using the participants' own VR equipment. We first conducted an online survey (N=\numberofparticipants) to understand HMD owners demographics, their VR setups and usage, and their willingness to participate in remote VR studies. Second, we analyzed existing app stores and social VR platforms to assess how suitable they would be to be utilized in remote VR studies, and what advantages and limitations they might bring. Third, we conducted two case studies to gather experiences from different approaches. Based on these contributions, we derived a framework, guiding researchers planning to conduct remote VR studies.

Our framework features four plausible paths to conduct remote VR studies. We learned that remote studies come with unique opportunities, challenges and considerations. The most defining factors in designing a suitable VR study for each purpose are 1) whether to utilize existing VR platforms or app stores, or to implement a standalone VR application, and 2) whether the study is run independently by participants, or guided by a remote experimenter.

We also identified best practices and lessons learned. Together with them and the framework, we believe that our work helps researchers to identify the best possible approach for their remote VR studies, and to anticipate potential pitfalls. As VR technologies keep advancing and more VR devices find their way into people's homes, we believe our work to be valuable to a growing number of researchers and practitioners.



\bibliographystyle{ACM-Reference-Format}
\bibliography{sample-base}

\appendix

\newpage



\end{document}